\documentclass[aps,pre,twocolumn,float]{revtex4}
\usepackage{amsmath,bm,epsfig}
\let \nn  \nonumber

\let\*\cdot
\def\<{\left\langle} \def\>{\right\rangle} \def\({\left(} \def\){\right)}
 \let\~\widetilde \let\^\widehat 

\def\be{\begin{equation}}\def\ee{\end{equation}}
\def\bea{\begin{eqnarray}}\def\eea{\end{eqnarray}}
\def\bse{\begin{subequations}}\def\ese{\end{subequations}}
\newcommand{\BE}[1]{\begin{equation}\label{#1}}
\newcommand{\BEA}[1]{\begin{eqnarray}\label{#1}}
\newcommand{\BSE}[1]{\begin{subequations}\label{#1}}

\let \nn  \nonumber
\usepackage{pstricks}
\usepackage{pst-node}
\usepackage[ansinew]{inputenc}
\usepackage{amssymb,amsmath}

\def\BSE{\begin{subequations}}\def\ESE{\end{subequations}}
\let \= \equiv

\def\a{\alpha}
\def\b{\beta}
\def\g{\gamma}

\def\o{\omega}

\def\be{\begin{equation}}       \def\ba{\begin{array}}

\def\ee{\end{equation}}         \def\ea{\end{array}}

\def\bea {\begin{eqnarray}}      \def\eea {\end{eqnarray}}

\def\bean{\begin{eqnarray*}}    \def\eean{\end{eqnarray*}}

\def\<{\langle} \def\({\left(}  \def\>{\rangle} \def\){\right)}

\newtheorem{exi}{Example}

\begin{document}

\title{Universal breaking point asymptotic for energy spectrum of Riemann waves\\ in weakly nonlinear non-dispersive media}
\author{Elena Kartashova$^{a}$, Efim Pelinovsky$^{a,b}$}
  \affiliation{$^a$Johannes Kepler University, Linz, Austria}
 \affiliation{$^b$ Institute of Applied Physics, Nizhny Novgorod, Russia}

   \begin{abstract}
In this Letter we study the form of the energy spectrum of Riemann waves in weakly nonlinear non-dispersive media.  For the simple wave equation with quadratic and cubic nonlinearity  we demonstrate that the deformation of a Riemann wave over time yields an
  exponential energy spectrum  which turns into a
power law asymptotic for high wave numbers with a slope of approximately  $-8/3$ at the last stage of evolution before
 the time of breaking  $T$. We argue, that this is the universal
asymptotic behavior of Riemann waves in any nonlinear non-dispersive medium at the point of breaking.  We also demonstrate that additional weak dispersion  or dissipation terms
(yielding the Korteweg-de Vries or Burgers equation respectively) do not change the universal asymptotic appearing at time $T$ though  no breaking occurs. Moreover, this universal asymptotic stays visible in the Fourier spectrum until times of order of $2T$ and more, for a wide range of smooth initial wave shapes.
The results reported in this Letter may be used in various non-dispersive media, e.g. magneto-hydro dynamics, physical oceanography, nonlinear acoustics.
\end{abstract}


\maketitle

\section{Introduction}
Weakly nonlinear wave systems fall into two categories -
dispersive and non-dispersive. In physical terms, for a
dispersive wave system group velocity changes
with frequency, for a non-dispersive wave system it does not.
Writing the wave
in the form of a sine wave $A \exp[i(kx-\o t)]$ we can
introduce the notion of
dispersion function $\o(k)$, $\o(k)$ being a real function, and define dispersive systems as
$\o_{k}^{''} \neq 0 $ and non-dispersive systems as $\o_{k}^{''} = 0 $, \cite{whith1}.

One of the most important characteristics of a wave system,
describing the  wave field as a whole,  is the distribution of
energy  over scales in Fourier space, i.e. the energy spectrum.

 For one initially excited \emph{dispersive} wave with a dispersion function of the form $\o(k)\sim k^{\b}$ it is known, that
the energy spectrum evolves from its initially exponential shape $e^{\g k}$, \cite{CUP,K12a},  into a power law $k^{a}$  \cite{K12a,ZLF92}), the point of transition from exponential to power law and the power law itself depending on the characteristic time and  the  dispersion function of the wave system, \cite{K12b,K13a}.

The power law "tail" of the energy spectra can be described deterministically, \cite{K12a} (in the wave systems with narrow frequency band excitation), or
statistically, \cite{ZLF92} (in the wave systems with distributed initial state).
Basic model for describing a dispersive weakly nonlinear system is the nonlinear Schr\"{o}dinger equation (NLS) and NLS-like equations modified to four and more nonlinear terms, e.g. \cite{DY79,Hog85}.

The class of the equations of the form
\begin{equation}\label{e1-trans}
u_t+(\a u +
\b u^2) u_x + \g u_{xxx}= 0
\end{equation}
known as Korteweg-de Vries-like models (KdV-like) widely used in soliton theory \cite{GPPT01}, dispersionless shock waves
\cite{GKE91,Lax91,Kam12} and soliton turbulence, e.g.
\cite{GZ99,PS06,EKPZ11}. Famous nonlinear Schr\"{o}dinger equation (NLS) and NLS-like models can be obtained from (\ref{e1-trans})
for narrow-band processes; accordingly the results obtained in  \cite{whith1,K12a,K12b,K13a,DY79,Hog85} are also valid for KdV-models.

The dynamics of a nonlinear \emph{non-dispersive} wave follow a simple model: one initially excited wave evolves into a  shock wave and finally breaks, \cite{whith1}.
The evolution of an unidirectional nonlinear wave before breaking is described by one nonlinear equation, sometimes called the simple
wave equation \cite{whith1,EFP88}, which reads
\begin{equation}\label{e1}
u_t+V(u) u_x = 0,
\end{equation}
where $u$ is the wave
function and $V(u)$ is a nonlinear local speed, $x$ is coordinate
and $t$ is time. For example, if we regard surface gravity waves
in shallow basin, $u$ is the water elevation and $V(u)
=3\sqrt{g(h+u)} - 2\sqrt{gh}$, where $g$ is acceleration due to
gravity, $h$ is unperturbed water depth \cite{St57,Did08}. A solution of (\ref{e1}) is called a Riemann wave.

Compare (\ref{e1-trans}) and
(\ref{e1}) we see that (\ref{e1}) can be regarded as  a dispersionless limit of (\ref{e1-trans}) and the nonlinearity $V(u)$ being
approximated by two terms of its Taylor expansion: $V(u)=\alpha u +
\beta u^2$ (any constant in the Taylor expansion of $V(u)$ can be
removed by an appropriate change of variables).

 The evolution of the wave field  after breaking depends
on the interplay of dispersion and dissipation.

In essentially
dissipative media, the shock wave persists and its amplitude spectrum
is known to have the high-frequency asymptotic $k^{-1}$,
\cite{GMS91,Kuz04}.

The dissipation - at least
weak dissipation - can be accounted for by including
a viscosity term $u_{xx}$ yielding, for quadratic nonlinear medium,
 the  Burgers equation,
 \cite{whith1}. It can then be  reduced to the linear
diffusion equation by the Hopf transformation and solved
explicitly. The Burgers equation with small
viscosity has been used for modeling one-dimensional turbulence  in
\cite{GSY83,Ru86,GMS91}.

In weakly dispersive media, modeled by KdV-like equations, solitons are developed; soliton turbulence  has been studied e.g. in \cite{GZ99,EKPZ11,PSSTEG13}. The energy
spectrum in this case and its dependence on
relevant  parameters such as initial wave shape have not yet been studied.

In this Letter
 we put forward a few important  novel questions about the  form of
energy spectra in a weakly nonlinear media:
Have energy spectra in dispersive and non-dispersive media similar shape?
What form of  energy spectra have Riemann waves? Does it depend on the type of nonlinearity? What form  has  spectrum asymptotic before breaking?  What is effect of dispersion? What is effect of dissipation? and others. To answer these questions we use (\ref{e1}) as the simplest possible mathematical model for studying non-dispersive media, Burgers equation for studying dissipative effects and  KdV equation for studying effect of dispersion.
\begin{figure}
\includegraphics[width=6cm]{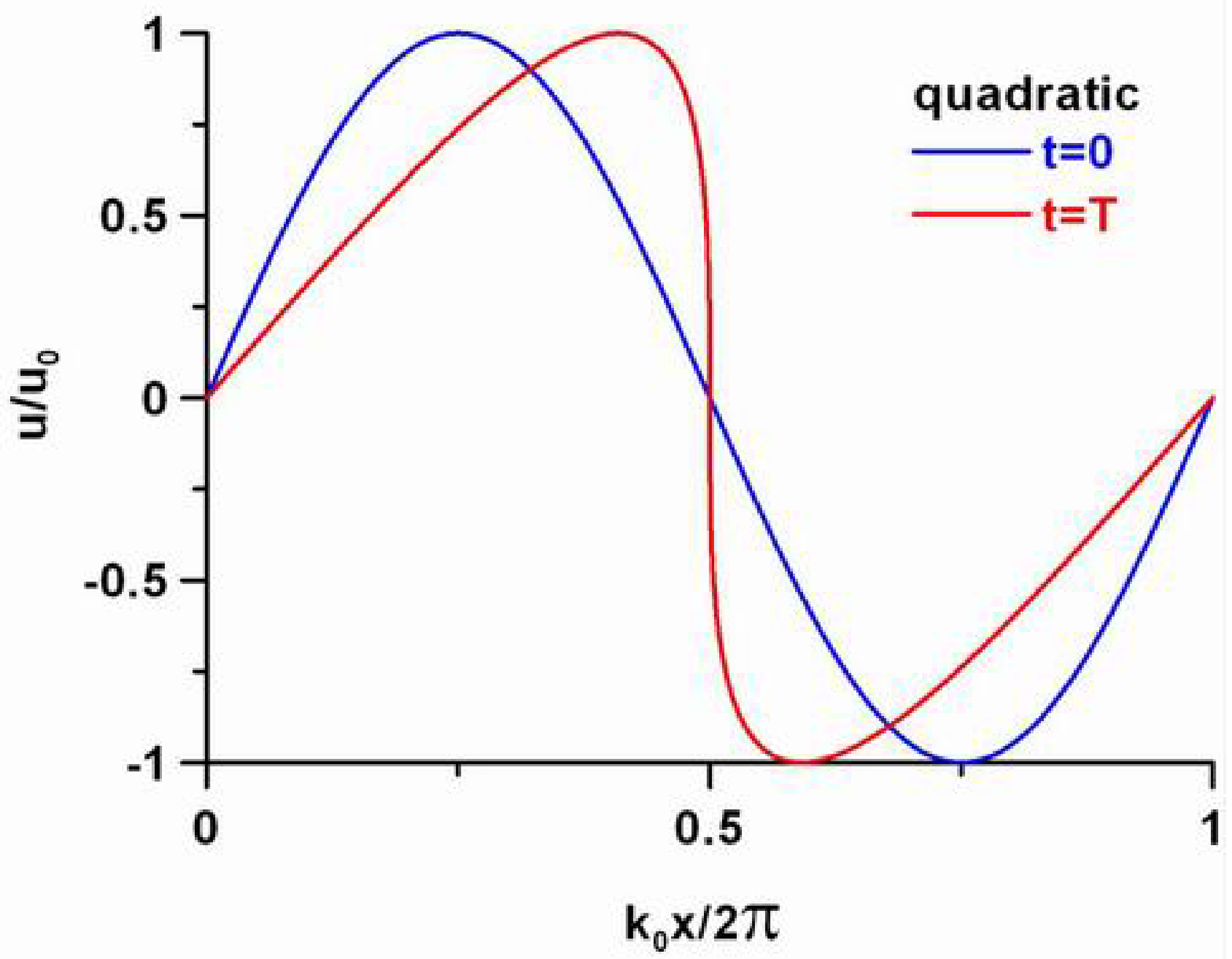}
\includegraphics[width=6cm]{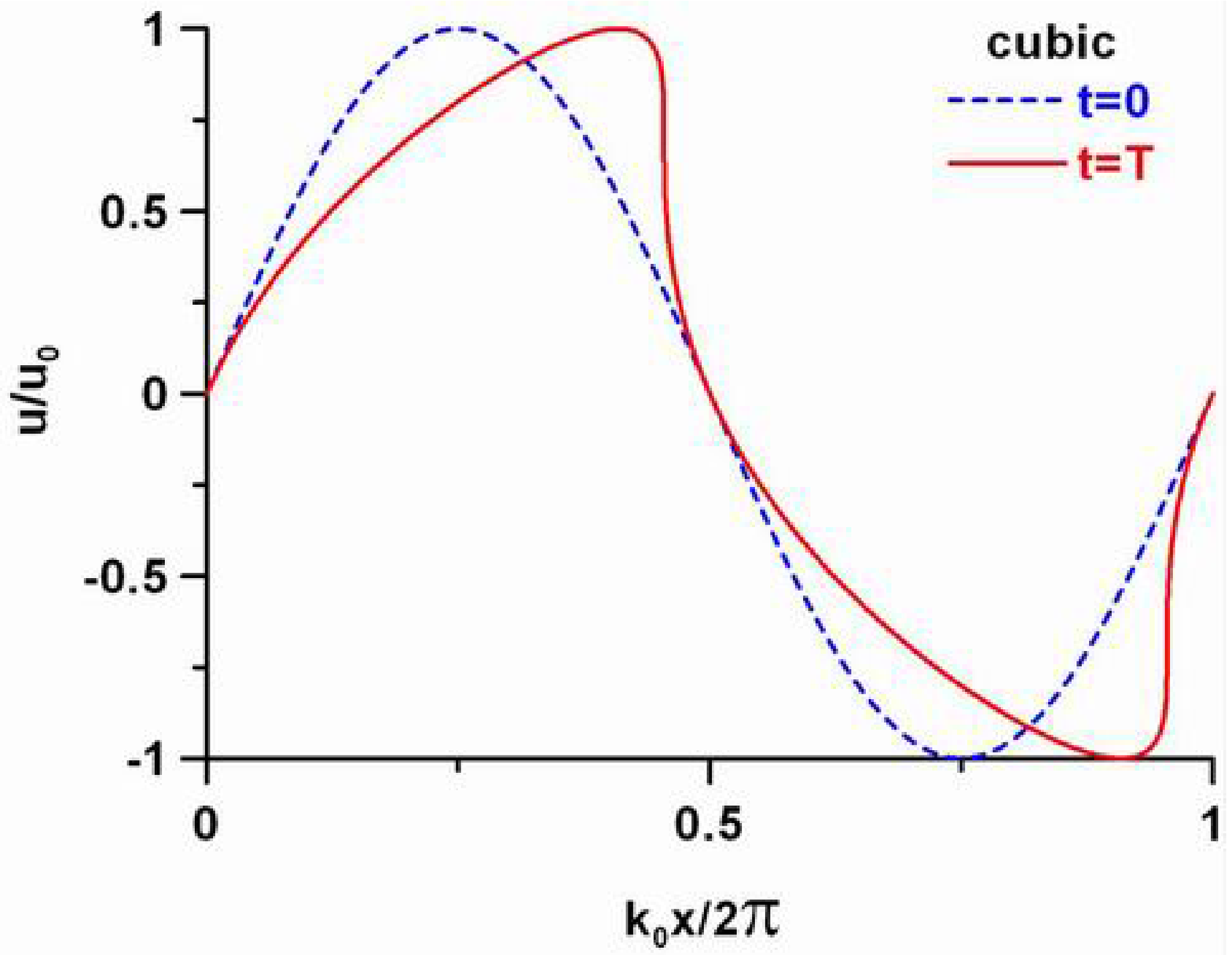}
\vskip -0.3cm
\caption{\label{f:12} Color online. Deformation of the Riemann wave $U(x)=U_0 \sin(k_0 x)$. \textbf{Upper panel}: Quadratic nonlinearity,  $\a\neq 0, \, \b=0$; the wave shape  is shown for $t=0$  and for
 $t = T =(\alpha k_0 U_0)^{-1}$. \textbf{Lower panel}: Cubic nonlinearity, $\a= 0, \, \b\neq0$; the wave shape is shown for $t=0$ and for
 $t = T =(\beta k_0 U_0^2)^{-1}$. In both panels, initial shape is shown in blue and the critical shape (at the moment of breaking) is shown in red.}
\vskip -0.5cm
\end{figure}

\section{Deformation of a Riemann wave before breaking}
The time evolution of a Riemann wave may be described as a
nonlinear deformation of the wave shape over time, which leads to
progressive growth of the wave steepness at one or more points of the wave period defined by wave height which eventually leads to wave breaking.
This may be seen the following way.

Solutions of (\ref{e1}) have the property, that any point of a given wave height $u(x)$ moves at constant speed. So we may write
\vskip -0.7cm
\be\label{e2}
u(x,t) = U[x - V(u) t],
\ee
where $U(x)$ is the initial
wave profile. Now computing
\vskip -0.5cm
 \be\label{e3}
u_x(x,t) = \frac{U_x}{ (1 + t V_x)},
\ee
we see that steepness $u_x$ is growing with time if
$V_x<0$.

In a general hyperbolic model,  if a wave reaches
 infinite steepness,  wave breaking will occur. This is called a gradient
catastrophe. The time of breaking $T$ may be computed as
\vskip -0.7cm
\be\label{e4}
T = 1/\max(-V_x).
\ee
 It follows from (\ref{e3}) that the
maximum value of wave steepness is proportional to \be\label{e14}
\max(u_x) \sim (T-t)^{-1} \ee

The result of the wave
deformation process depends on the  local speed $V(u)$ which in turn depends on the form of nonlinearity. This is
illustrated in Fig. \ref{f:12} for quadratic and cubic nonlinearity. In both panels, the
initial wave  has the shape of a sine
\begin{equation}\label{e5}
U(x)=U_0 \sin(k_0 x).
\end{equation}

In the quadratic  nonlinear media only one shock is formed within a wave period and media breaking
occurs for wave height $u/U_0 =0$ at the moment of time $T=(\alpha k_0
U_0)^{-1}$.

In the cubic nonlinear media two shocks within the wave period are formed, with
opposite sign of slope, and breaking occurs for wave heights $u/U_0 = \pm
\sqrt{2}/2$ simultaneously at the moment of time $T=(\beta k_0 U_0^2)^{-1}$.
\begin{figure}
\includegraphics[width=6cm]{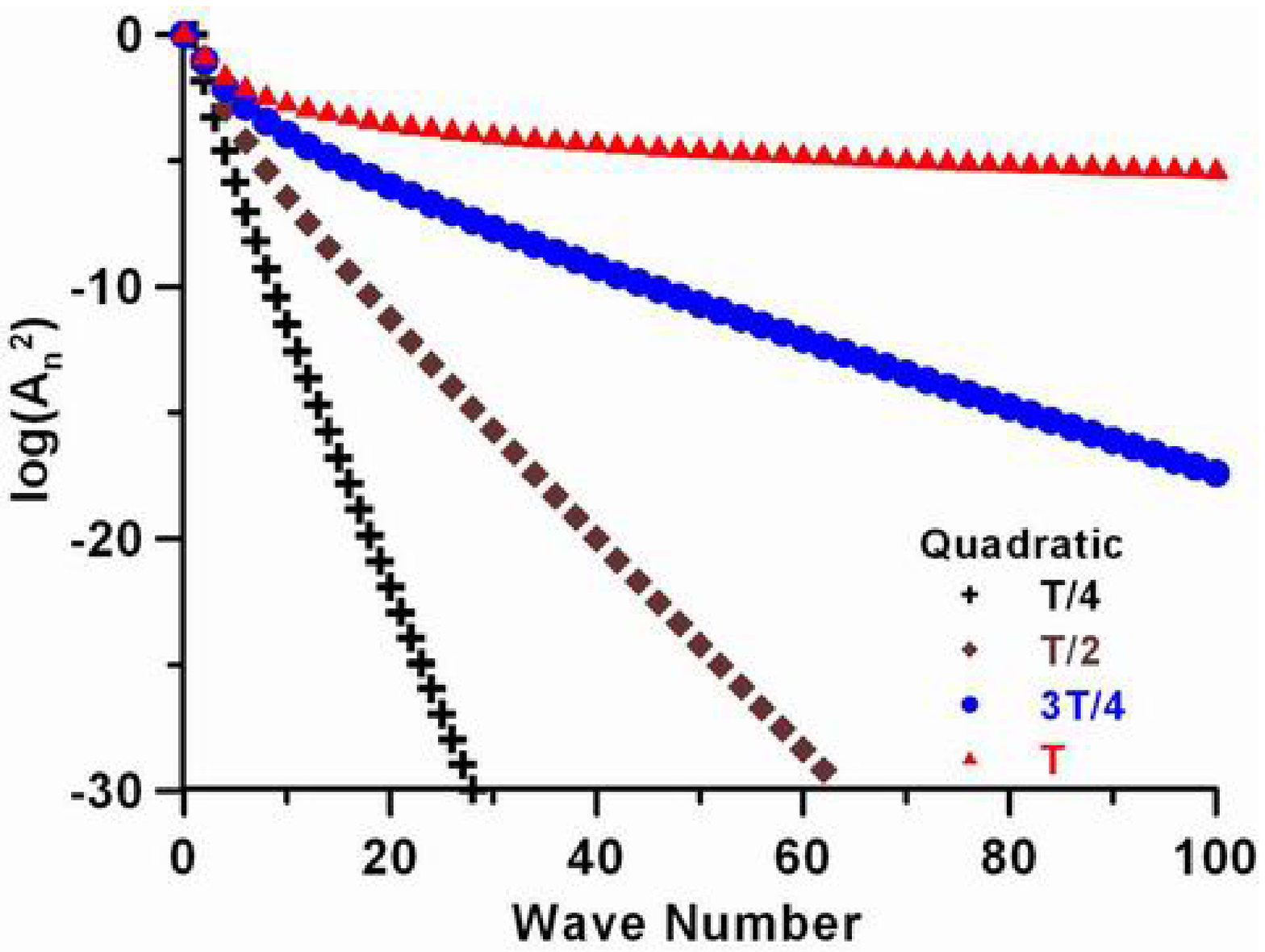}
\includegraphics[width=6cm]{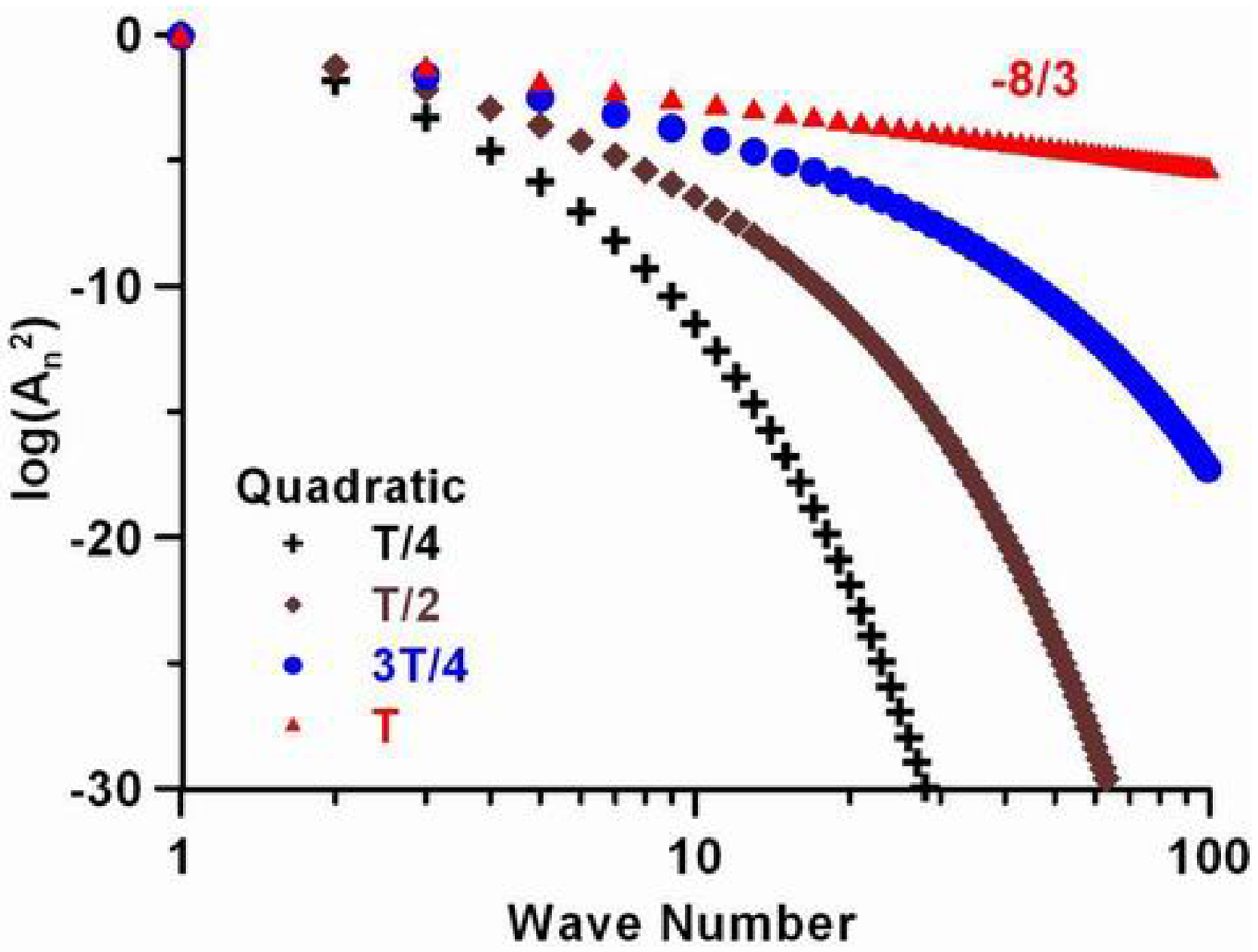}
\caption{\label{f:34} Color online. Energy spectrum of Riemann waves in quadratic nonlinear medium at
different moments of time $t= 0.25T, 0.5T, 0.75T, T$, shown in  semi-logarithmic coordinates (\textbf{upper panel}) and in logarithmic coordinates (\textbf{lower panel})}
\end{figure}
\begin{figure}
\includegraphics[width=6cm]{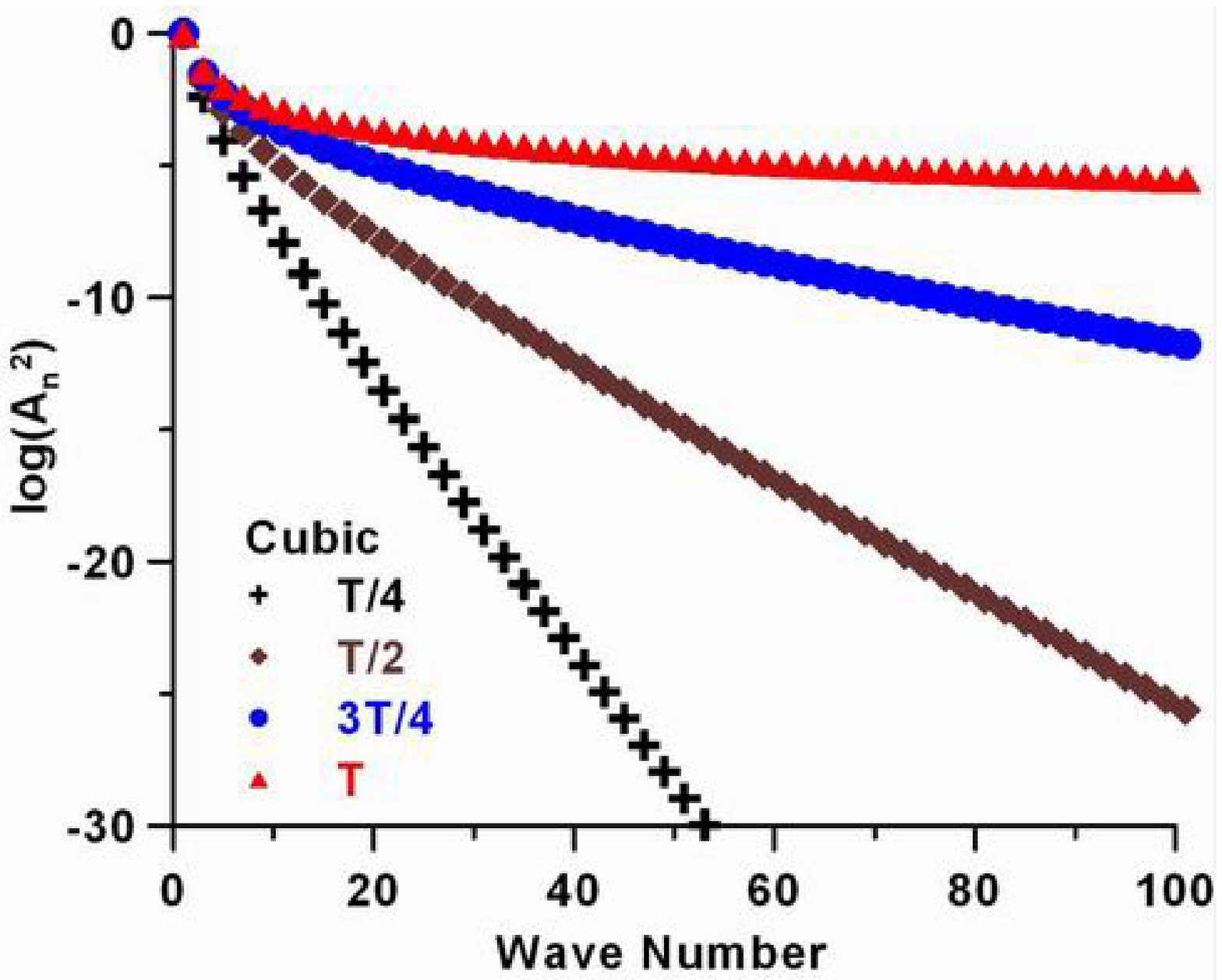}
\includegraphics[width=6cm]{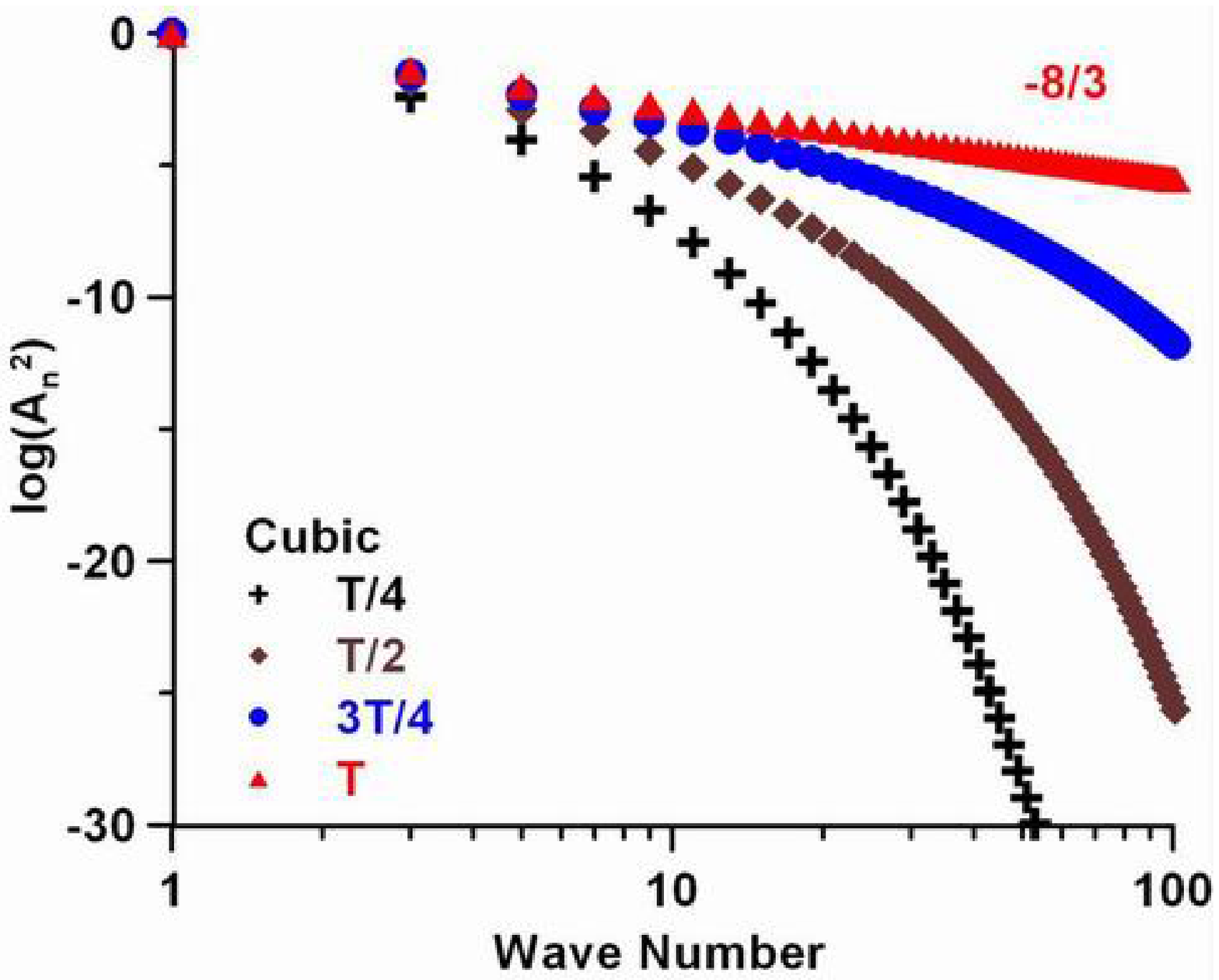}
\caption{\label{f:56} Color online. Energy spectrum of Riemann waves in cubic nonlinear medium at
different moments of time $t= 0.25T, 0.5T, 0.75T, T$, shown in  semi-logarithmic coordinates (\textbf{upper panel}) and in logarithmic coordinates (\textbf{lower panel})}
\end{figure}
\section{Fourier spectra of Riemann waves}
The spatial spectrum of a  wave $u(x,t)$ is defined as
\begin{equation}\label{e6}
S(k,t)=\int_{- \infty}^{+ \infty} u(x,t) \exp(-ikx)dx.
\end{equation}
Next we compute the spatial spectrum of a Riemann wave in explicit form.

Having in mind (\ref{e3}) we perform the change of variables $X = x - V(u) t$ yielding
\begin{equation}\label{e7}
dx = \frac{dX}{1-t V_x}.
\end{equation}
Now we can rewrite (\ref{e6}) as
\begin{equation}\label{e8}
S(k,t) = \int_{- \infty}^{+ \infty} (1 + V_X) U(X)
\exp[-ik(X+Vt)] dX.
\end{equation}
By simple manipulation, this is transformed to
\begin{equation}\label{e9}
S(k,t)=(-ik)^{-1} \int_{- \infty}^{+ \infty} U_X \exp[-ik(X+tV)]
dX.
\end{equation}
For one initially exited sine wave as given by
(\ref{e5}), the integral in (\ref{e9}) may be computed analytically, both for
 quadratic and  cubic nonlinearity, \cite{Peli76}.

Using (\ref{e1-trans}) with $\a> 0,\,\beta = 0$  we get \emph{quadratic nonlinearity}, and the wave field may be
represented by the Bessel-Fubini series well-known in nonlinear
acoustics, \cite{Peli76}:
\vskip -0.8cm
\bea
u(x,t)=U_0 \sum_{n = 1}^{\infty} A_n (\tau) \sin [nk_0 x],\label{e10}\\
\mbox{where} \qquad
A_n (\tau) = \frac{2 (-1)^{n+1}}{n \tau} J_n (n\tau),   \quad
\tau = t/T,\label{e11}
\eea
with $T = (\alpha U_0 k_0)^{-1}$ being the time of breaking, and $J_n
(z)$ being the Bessel function.

Fig.\ref{f:34} depicts the time evolution of the energy spectrum given as values $A_n^2$ as a function of wave number $k_n$ showing clearly:
Up to the moment of time
$t=3T/4$ the spectrum in semi-logarithmic coordinates (upper
panel) has the form of a straight line, which means it is
exponential in linear coordinates. As time grows from $t=3T/4$  to the moment of
breaking $t=T$, a power law spectrum is formed, with a slope of
$2.67$, which is close to  8/3.
\begin{figure}
\includegraphics[width=6cm]{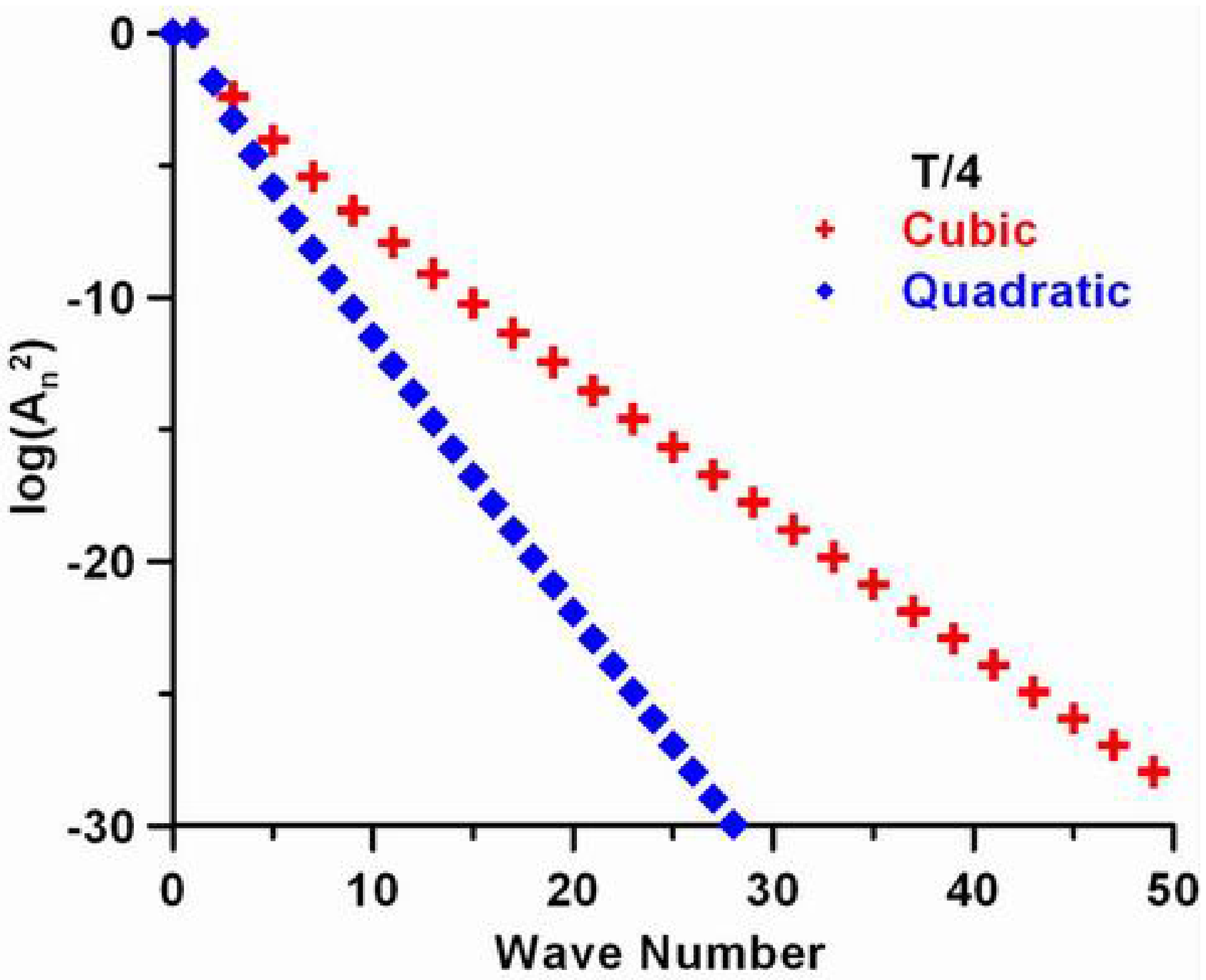}
\includegraphics[width=6cm]{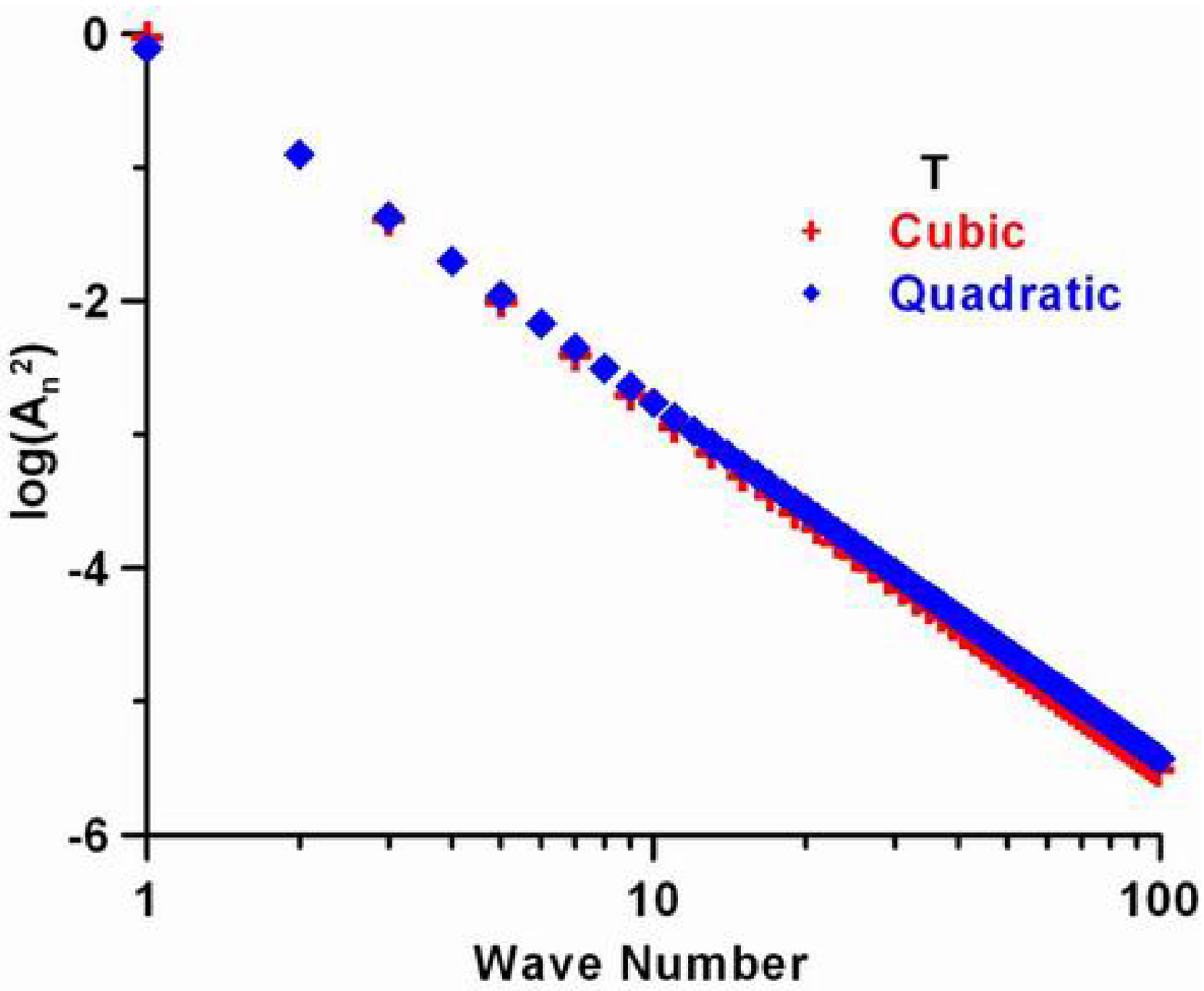}
\caption{\label{f:78} Color online. \textbf{Upper panel}: Exponential "part" of spectrum in  quadratic and  cubic media, shown by blue crosses and red circles respectively. \textbf{Lower panel}: Power law "part" of spectrum in  quadratic and  cubic media, color scheme as above. }
\vskip -0.3cm
\end{figure}
Using (\ref{e1-trans}) with $\a= 0,\,\beta >0$
 we get \emph{cubic nonlinearity}. The wave field may be
represented by a
Bessel-Fubuni-like series, \cite{Peli76},  of the form \\

$\,
u(x,t)=$ \bea\label{e12} &&U_0 \sum_{n = 0}^{\infty}
\frac{1}{2n+1} \large\{J_n
[(n+1/2)\tau] \sin [(2n+1)(k_0 x - \tau/2] \nonumber \\
&& + J_{n+1} [(n+1/2)
\tau] \cos [(2n+1)(k_0 x - \tau/2] \large\},
\eea
where again $\tau = t/T$ and the time of breaking  now is given by  $T = (\beta k_0
U_0^2)^{-1}$. Unlike the quadratic case where all Fourier harmonics are in phase in the cubic case we also have a phase shift of the Fourier harmonics, and
the normalized amplitudes are
\begin{equation}\label{e13}
A_n = \sqrt{J_n^2 [(n+1/2)\tau] + J_{n+1}^2 [(n+1/2)\tau]}.
\end{equation}

The energy spectrum is shown in Fig.\ref{f:56}. As in the case of a quadratic
nonlinearity,  it has exponential shape  for  small times (upper panel)
and turns into a power law  when  time
approaches the moment of breaking, $t \rightarrow T$. The slope
of the power law is  $2.67$ and again close  to  8/3.

It is clearly seen from Fig.\ref{f:78} (upper panel) that
the energy spectrum has the
exponential part, both for
quadratic and cubic nonlinear media.   The power
law part of the energy spectrum is shown in Fig.\ref{f:78}, lower panel, for quadratic and cubic nonlinear media; the shape of the energy spectrum is practically undistinguishable for these two cases.

In the next two sections we present the results of numerical simulations which show the effect of dispersion (Sec.\ref{s:dispersion}) and dissipation (Sec.\ref{s:dissipation}) on the formation of the universal asymptotic in the energy spectrum of Riemann waves. As a reference point we regard a particular case of (\ref{e1-trans}) with $\a=6, \, \b=0, \, \g=0$:
\vskip -0.5cm
\be\label{KdV-noD}
u_t+6 u u_x = 0.
\ee
\vskip -0.2cm
This is the dispersionless Korteweg-de Vries equation (KdV) in canonic form. All numerical simulations presented in the next sections have been performed with  an initially  sinusoidal wave of the form (\ref{e5}), a tank length $L=200$ and initial disturbance $U=U_0\sin{k_0x}$ such that the wave length equals the length of the tank, which means $k_0=0.0314$. The breaking time is $T= 26.5$.

Numerical simulations with other forms of smooth and physically relevant initial conditions - e.g. $U =U_0\exp{(- x^{2k_0})}$ - have also been conducted; the results are quite similar to those  with sinusoidal initial condition, and are not shown in the text below.

\section{Effect of dispersion}\label{s:dispersion}
Adding to (\ref{KdV-noD})  a dispersive term of the form $u_{xxx}$ we get
\be\label{KdV}
u_t+6 u u_x + u_{xxx}= 0,
\ee
 the classical KdV equation. The transformation of a sinusoidal wave into a group of solitons or cnoidal waves  is a classical problem which has been studied in the frame of the KdV equation,  with  main interest on the evolution of the wave shape  in physical space. Our  interest is to examine the form of the energy spectrum in Fourier space and if something which resembles the universal asymptotic $-8/3$ may be observed under the influence of weak dispersion.

What happens in this case depends on the Ursell number, which is the ratio of the nonlinear term to the dispersive term:
\be \label{Ursell}
Ur = \max_x (u u_x)/\max_x (u_{xxx}).
\ee
For shallow water waves the Ursell number can be computed as $Ur=A\cdot L^2/h^3$ where $A$ and  $L$ are the wave amplitude and wave length respectively,  and $h$ is the depth of the water layer, $h \ll L$. Accordingly, the Ursell parameter changes with the wave length $L$.

\begin{figure}
\includegraphics[width=7cm]{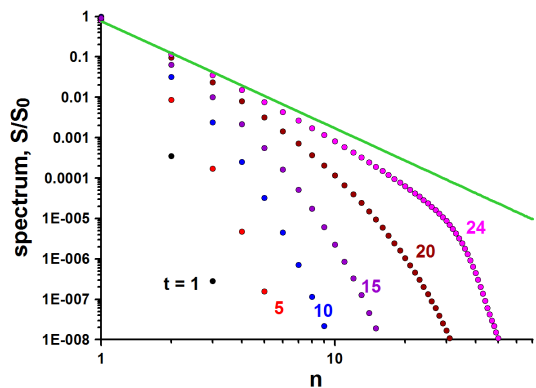}
\caption{\label{f:1} Color online. Formation of the breaking time asymptotic $k^{-8/3}$ (shown by bold  green line) in KdV equation (\ref{KdV}) with initial conditions (\ref{e5}). Each curve represents the spectrum at a certain  moment of time $t=5, 10, 15, 20, 25 < T=26.5$ (a moment of time is shown as a number written on the left of each curve). Axes $n$ (wave number) and $S/S_0$ are shown in logarithmic coordinates.}
\end{figure}

For our numerical simulation, with $L$ chosen as $L=200$, we get $Ur=L^2=40000$ and the effect of dispersion should be small. Taking again initial conditions of the form (\ref{e5}), we observe the same time evolution as in  the dispersionless KdV (\ref{KdV-noD}) at different times $t$ smaller than breaking time $T$; the results of our simulation are shown in Fig.\ref{f:1}.

\begin{figure}
\includegraphics[width=6cm]{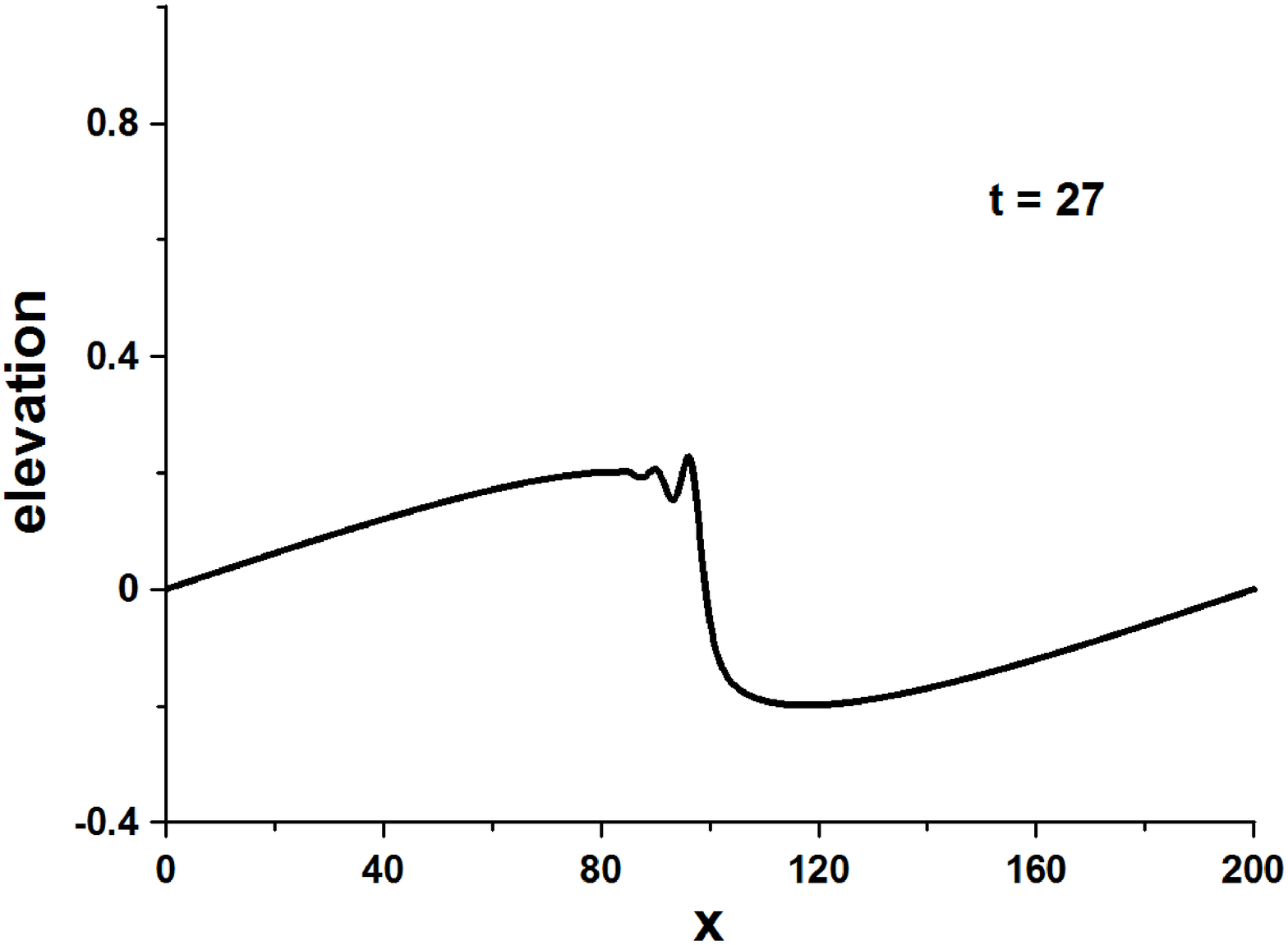}
\includegraphics[width=6cm]{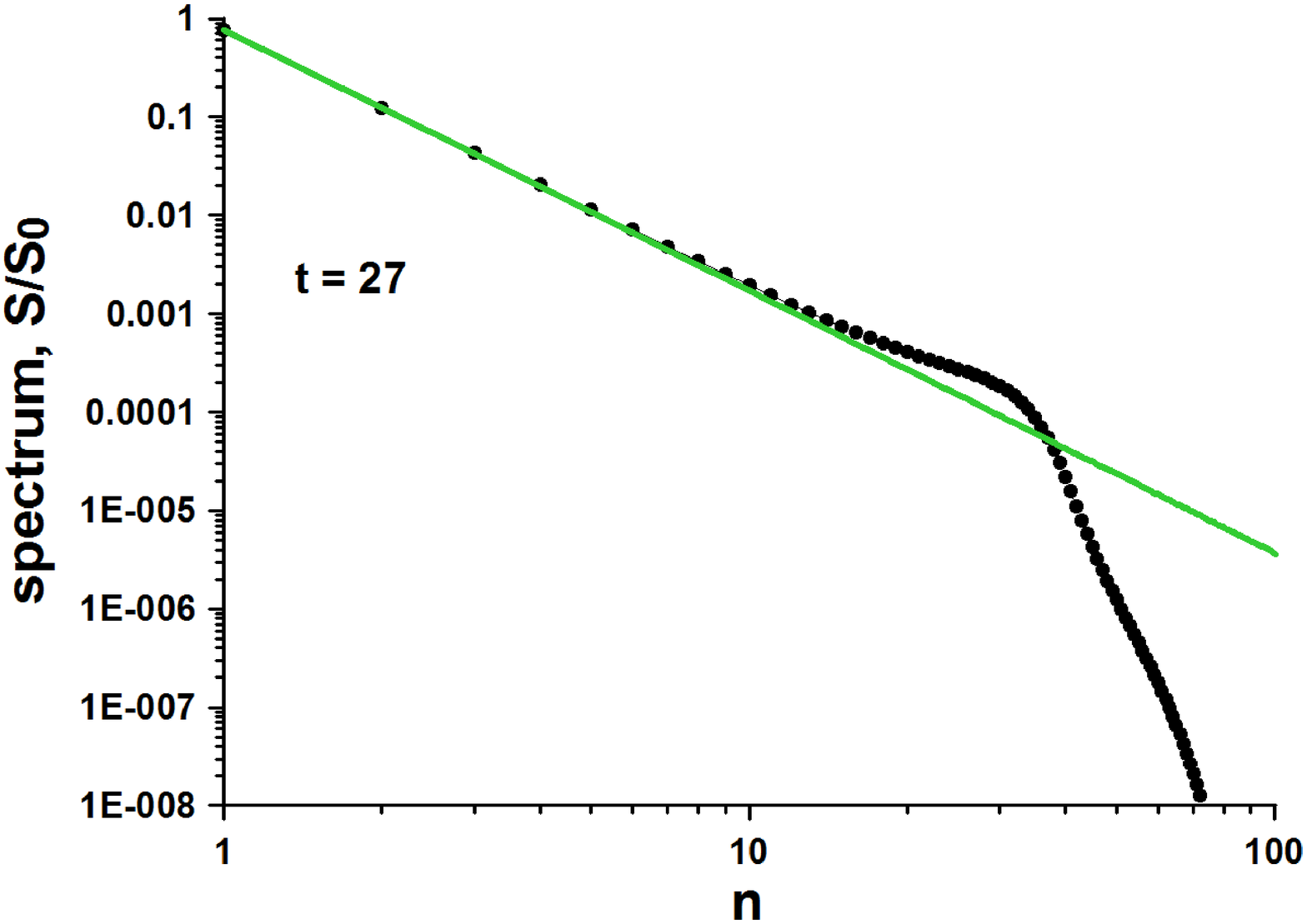}
\caption{\label{f:2} Color online. Same as in Fig.\ref{f:1} but for the moment of time $t=27$. }
\end{figure}

When $t$ goes beyond breaking time $T=25.5$, the Fourier spectrum changes rapidly.  Already at the  moment of time $t=27$ the formation of an isolated peak in the energy spectrum is visible as shown in Fig.\ref{f:2}, lower panel; the shape of the surface elevation corresponding to this is shown in Fig.\ref{f:2}, upper panel.
 \begin{figure}
\includegraphics[width=6cm]{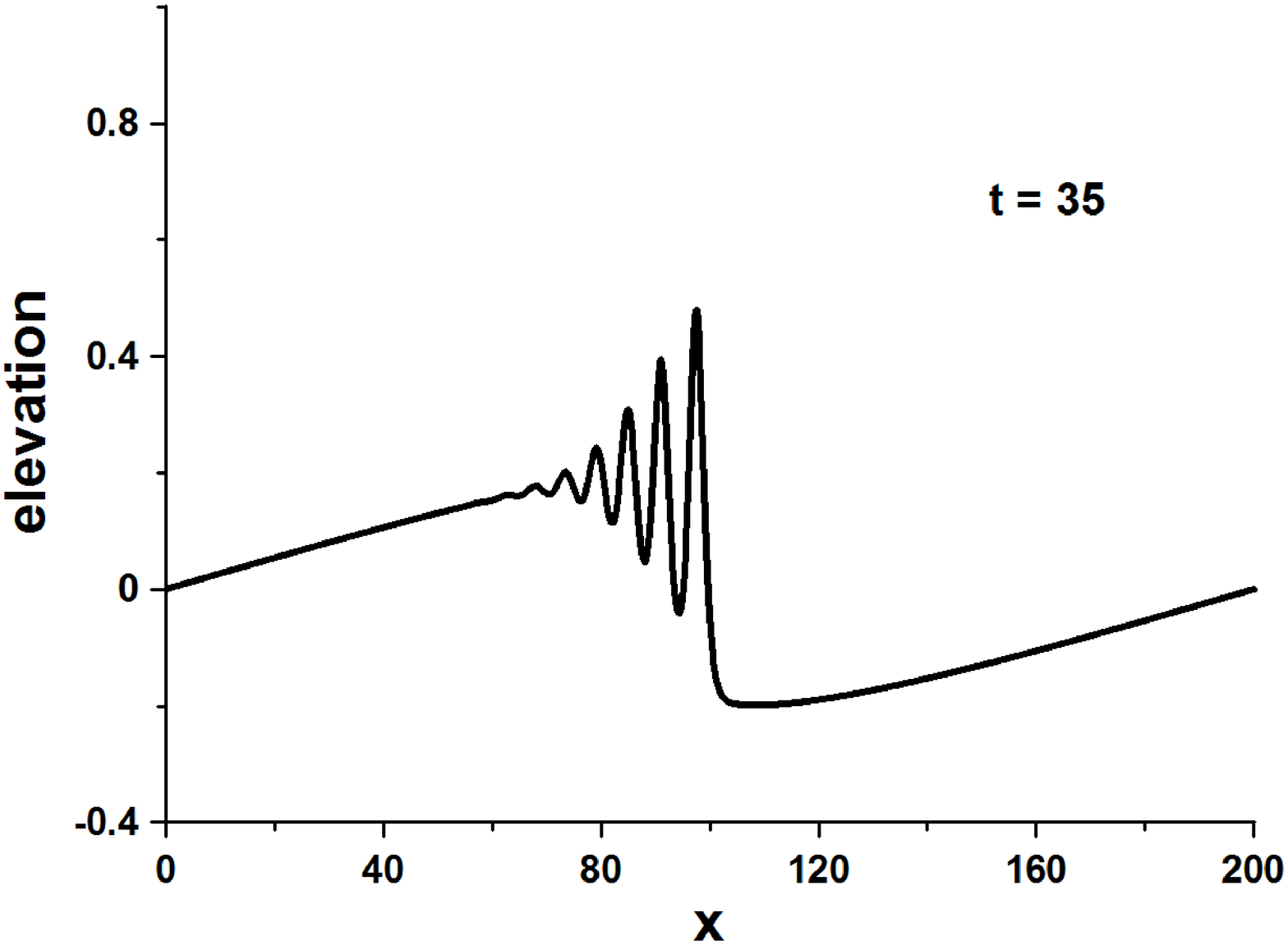}
\includegraphics[width=6cm]{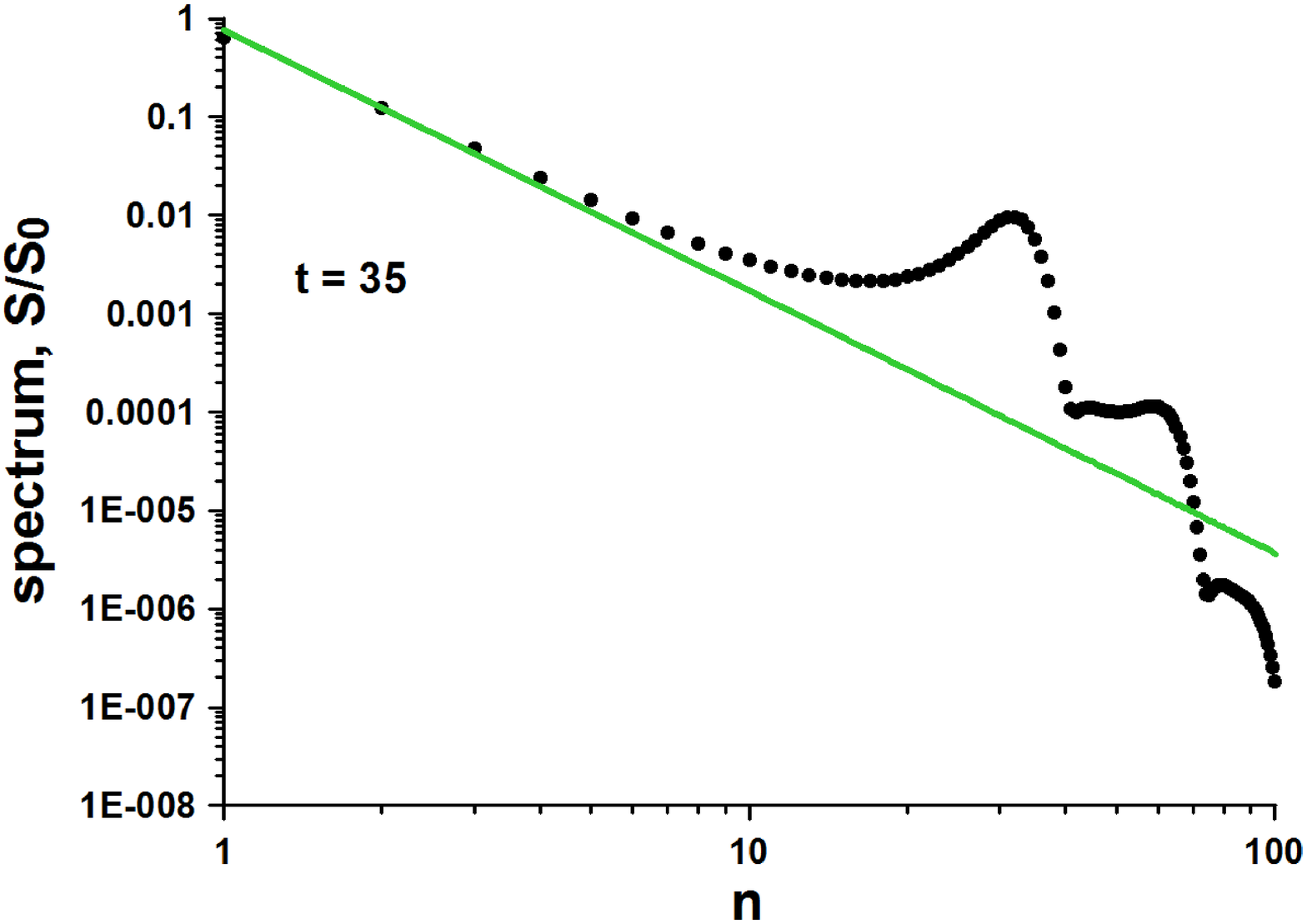}
\caption{\label{f:3} Color online. Same as in Fig.\ref{f:1} but for the moment of time $t=35$. }
\end{figure}
 \begin{figure}
\includegraphics[width=6cm]{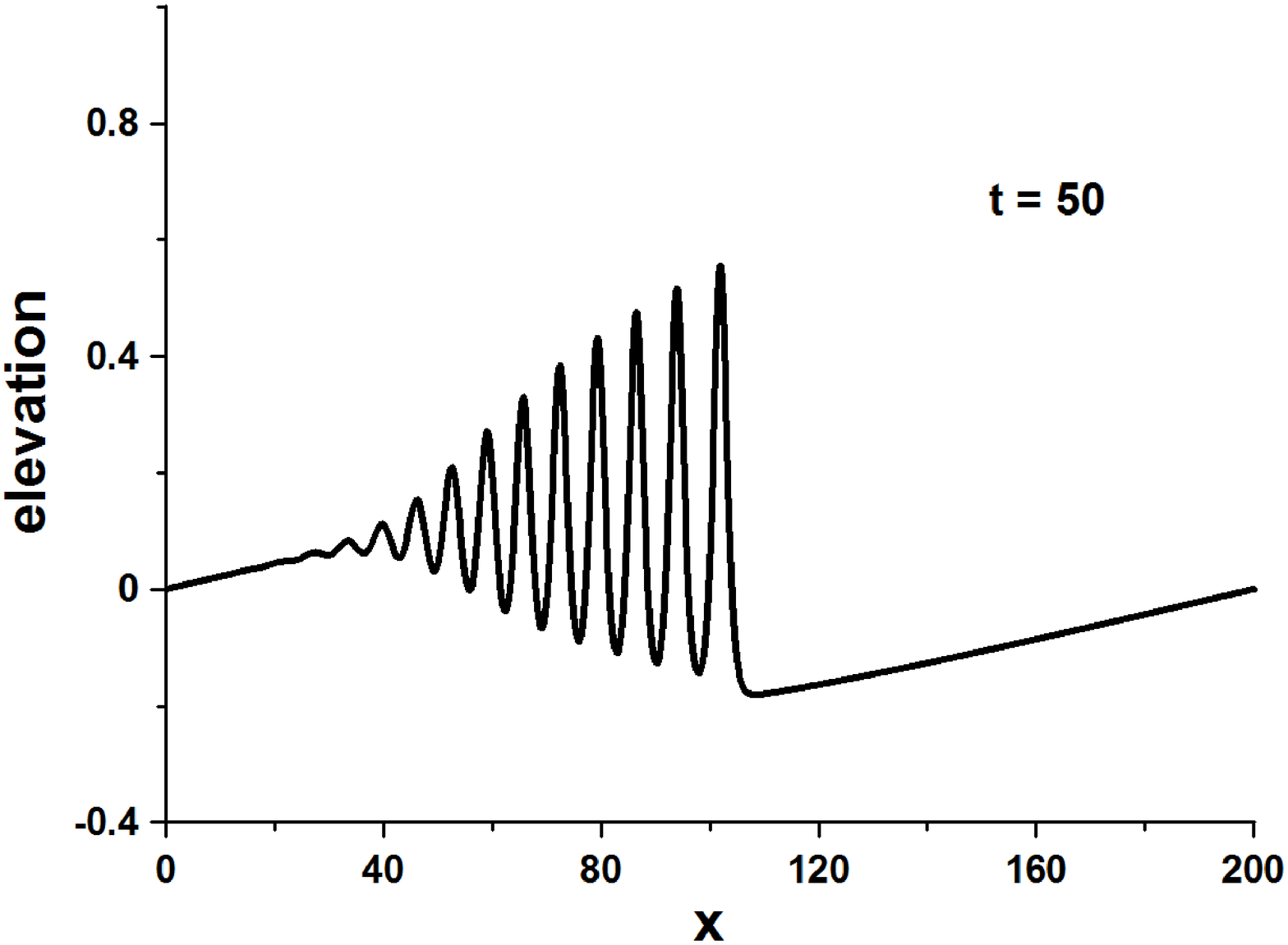}
\includegraphics[width=6cm]{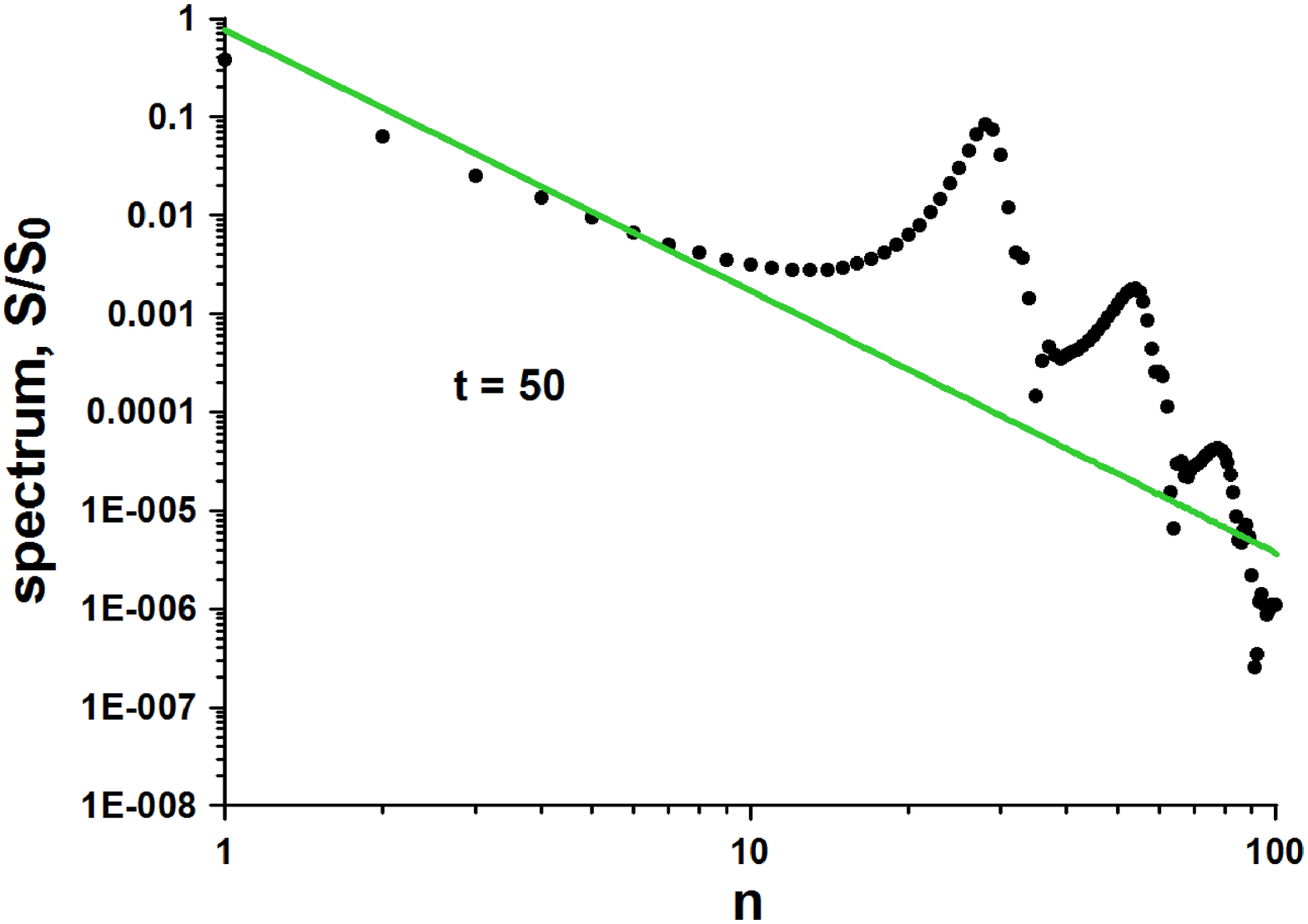}
\caption{\label{f:4} Color online. Same as in Fig.\ref{f:1} but for the moment of time $t=50$. }
\end{figure}
 \begin{figure}
\includegraphics[width=6cm]{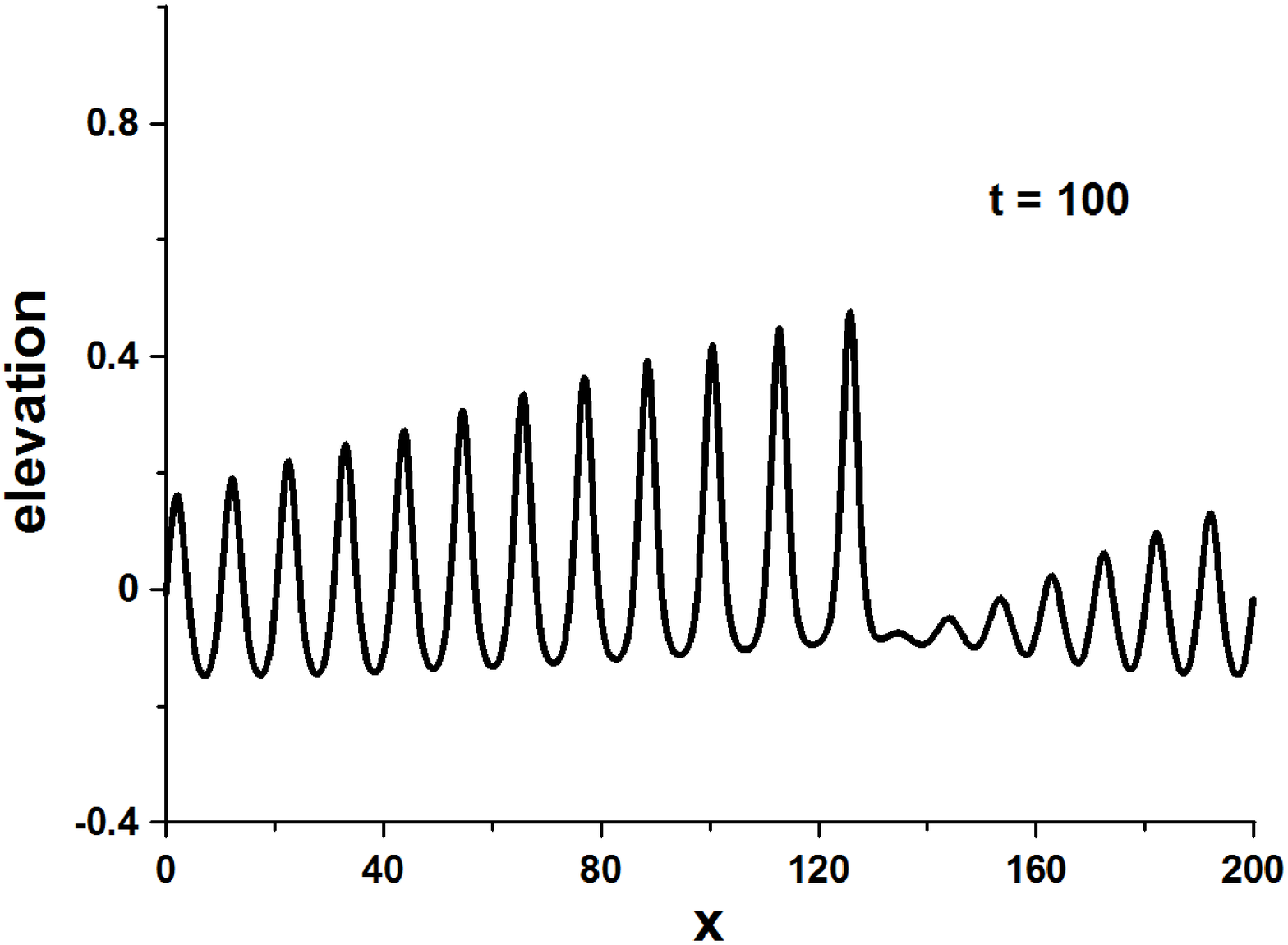}
\includegraphics[width=6cm]{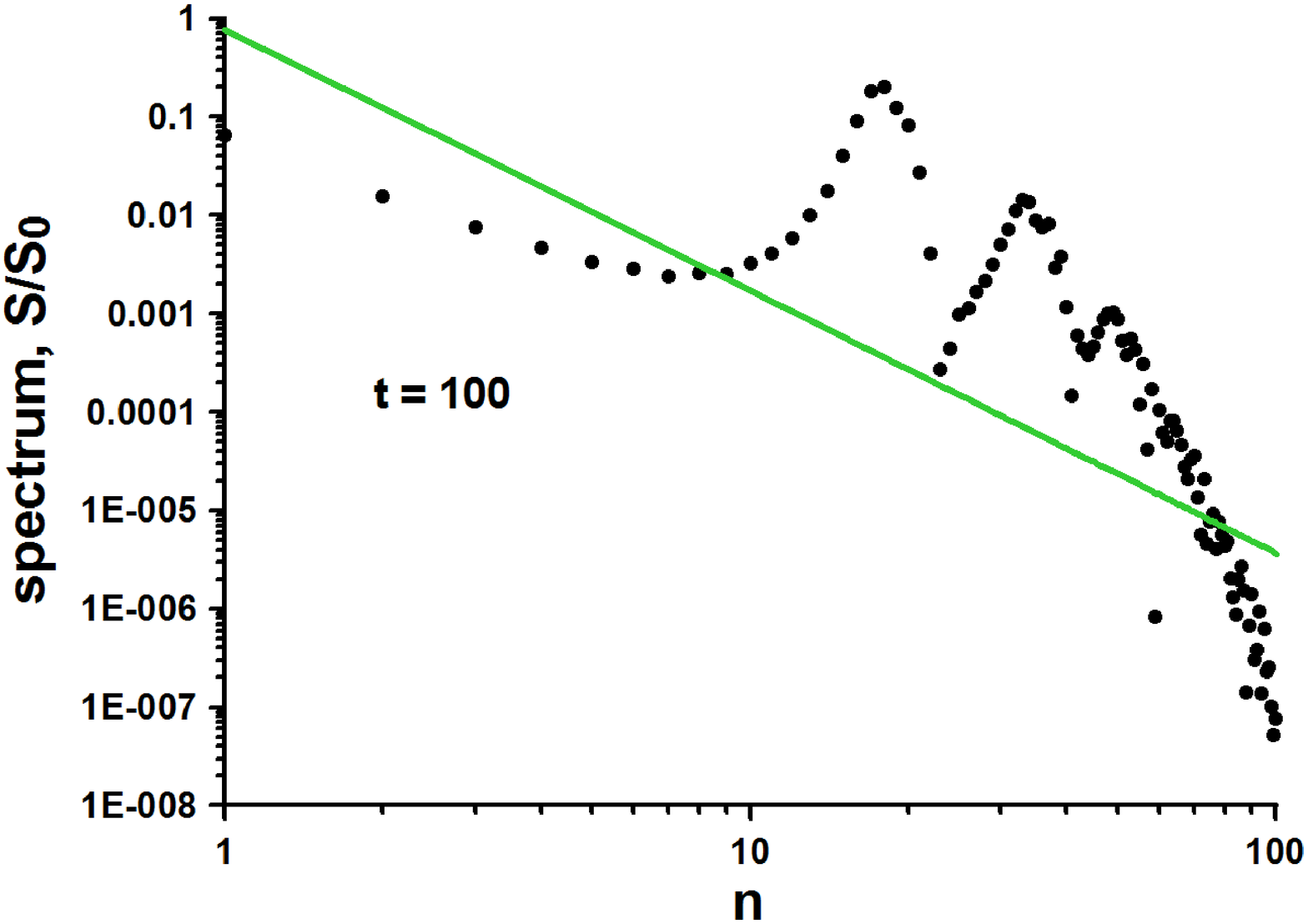}
\caption{\label{f:5} Color online. Same as in Fig.\ref{f:1} but for the moment of time $t=100$. }
\end{figure}

 The further development is characterized by the formation of soliton-like disturbances in physical space (upper panels, Figs.\ref{f:3}-\ref{f:5}) leading to the appearance of narrow spectral peaks in the high-frequency range in Fourier space (lower panels, Figs.\ref{f:3}-\ref{f:5}). The first spectral peak appears in the Fourier harmonic with $n=34$ at the moment of time $t = 27$, as a soliton with soliton width $6$  is the 34 th harmonic of a wave with initial wavelength $200$ ( $200/6 \approx 34$). During time evolution this peak is downshifted to the left, so that at  moment of time $t=100$ the maximum of energy is at the harmonic with $n=18$ which corresponds to a soliton width of $\approx 10.5$.

 At a later time moment, new soliton-like disturbances become detectable in the Fourier spectrum, with essentially  the same behavior. Thus the second peak appears at the harmonic with $n=60$ and moves to the left, finally arriving at the harmonic with $n=40$; the soliton amplitude is growing and  the soliton width is decreasing.

 Thus, the energy spectrum consists of two substantially different parts: the universal power law asymptotic $k^{-8/3}$ at the initial stage after breaking and an exponential asymptotic of the newly appearing solitons (not shown in the figures).

 The universal breaking time asymptotic $k^{-8/3}$ stays unchanged up to the moment of time $t \approx 2 T$.
 After that,  the power law part of the spectrum becomes shorter in Fourier space applying only to low frequencies and energy decays more slowly compared to the exponent $-8/3$; the "soliton"-part of the spectrum takes over from the right.
 For the chosen $Ur$ (which is the relation between nonlinearity and dispersion) and time $t \approx 4 T$, three peaks in Fourier space are clearly visible in the figures. For bigger times the number of solitons in physical space is growing and they do not describe the front structure anymore as the width of the front becomes comparable to the initial wave length $L$.

\section{Effect of dissipation}\label{s:dissipation}
The usual way of modeling dissipation is to add to (\ref{KdV-noD}) a term for dissipation  $-\nu u_{xx}$ with a small coefficient of viscosity $\nu>0.$  The resulting equation
\be\label{Burgers}
u_t+6 u u_x - \nu u_{xx}= 0
\ee
is the well known Burgers equation, which has a  long time asymptotic $k^{-2}$ for the energy spectrum, \cite{GMS91}.

As it is  shown above, if $\nu=0,$ the breaking time asymptotic is $k^{-8/3}$. Our  interest is to examine the form of the energy spectrum in Fourier space and if something which resembles the universal asymptotic $-8/3$ may be observed under the influence of dissipation.

\begin{figure}
\includegraphics[width=6cm]{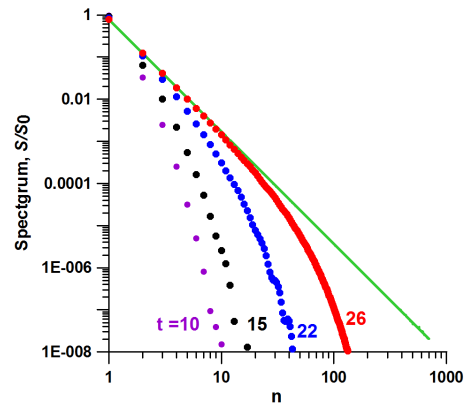}
\caption{\label{f:6} Color online. Formation of the breaking time asymptotic $k^{-8/3}$ (shown by bold  green line) in Burgers equation (\ref{Burgers}) with initial conditions (\ref{e5}). Each curve represents the spectrum at a certain  moment of time $t=10,15,22,26 < T$ (a moment of time is shown as a number written on the left of each curve). Axes $n$ (wave number) and $S/S_0$ are shown in logarithmic coordinates.}
\end{figure}

With this aim we study the Burgers equation numerically, with initial condition (\ref{e5}) as above and the
viscosity coefficient taken as $\nu=0.1$. In Fig.\ref{f:6} the formation of the breaking time asymptotic $k^{-8/3}$ is shown, depicting the spectrum at $t=10,15,22,26$ before breaking time $T=26,5.$
 \begin{figure}
\includegraphics[width=6cm]{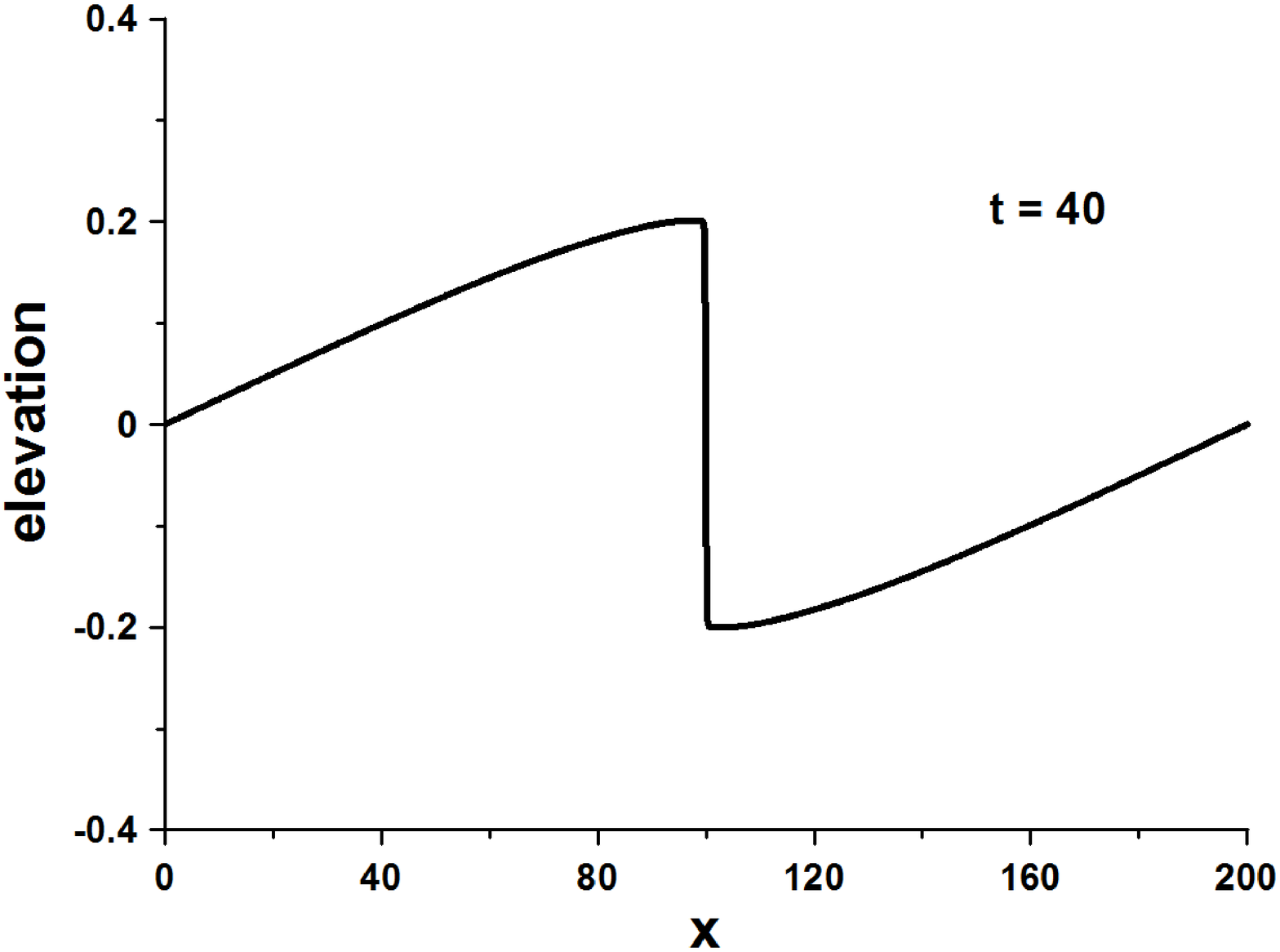}
\includegraphics[width=6cm]{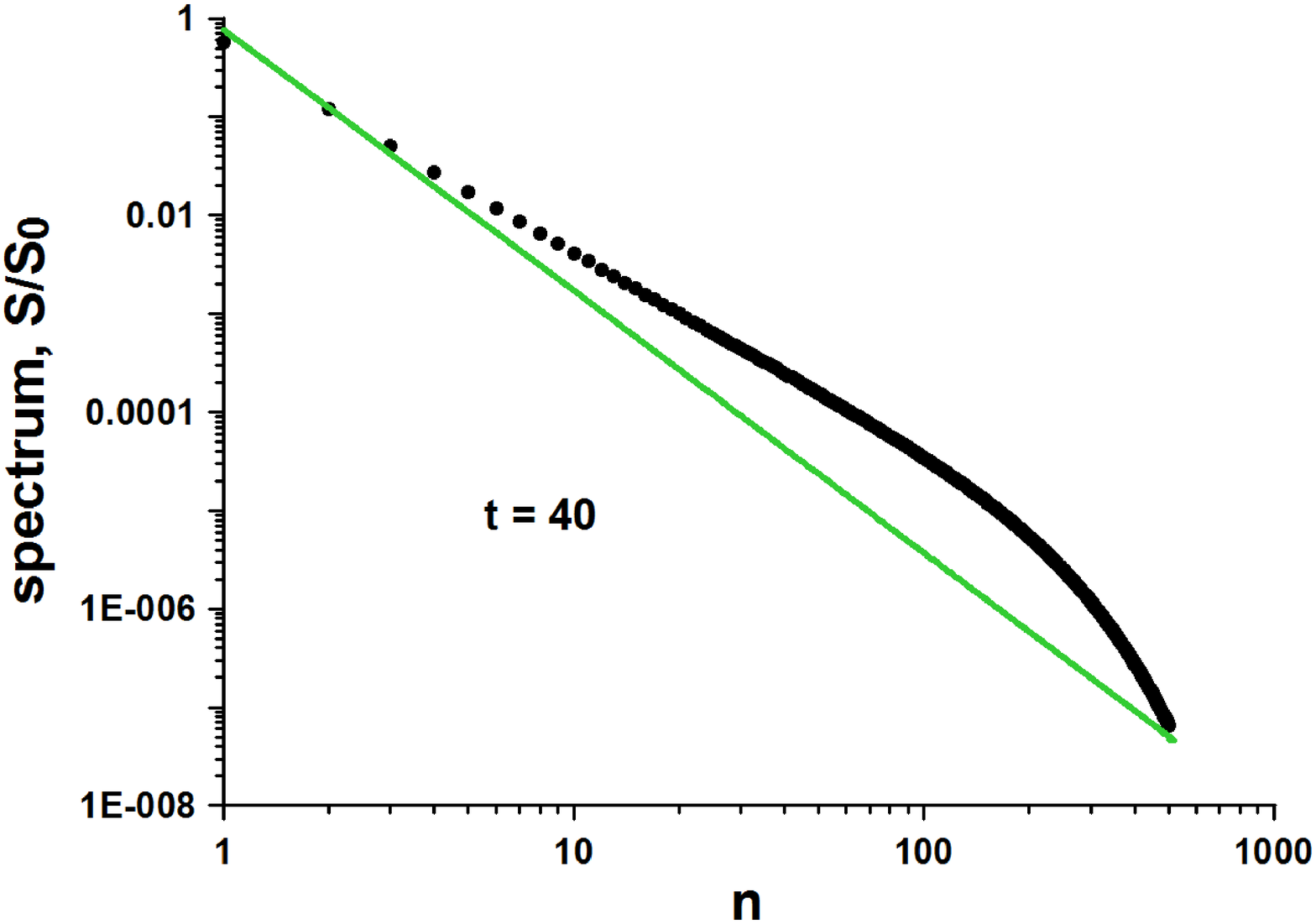}
\caption{\label{f:8} Color online. Spectrum form for time after breaking time $t=40>T.$ The wave shape in physical space is shown  on in the upper panel, and corresponding Fourier spectrum - in the lower panel.}
\end{figure}

Beginning with  $t=27 > T,$ the appearance of a new asymptotic which decays more slowly than $k^{-8/3}$ becomes evident; it is shown in Fig.\ref{f:8}. With increase in time the wave shape turns into a triangle and its spectrum $k^{-2}$ can be computed analytically, \cite{GMS91}. For  $t > 5T$ the initial wave is damped significantly and the spectrum becomes narrow again (not shown in figures).

\section{Discussion and Conclusions}
The results presented in this Letter can be summarized as follows.

-- We have found approximately the same power law of $\approx k^{-8/3}$
 for the energy spectrum of Riemann waves
before breaking in a vicinity of the breaking point, in quadratic as well as in cubic media. As the
quadratic and cubic nonlinearity are described by the first and
second term of the Taylor expansion, this is an indication, but not a proof,
 that $k^{-8/3}$ is the universal power law for Riemann
waves with any nonlinearity.

-- This may be re-formulated in the following way: any point of singularity in the (initially smooth) wave profile of a Riemann wave before breaking may be described by a power law of $x^{1/3}$. Indeed, the Fourier spectrum of the power function $x^q$ is $k^{-(1+q)}$, and the corresponding energy spectrum is $k^{-2(1+q)}$. Taking  $q=1/3$ we get our energy spectrum of  $k^{-8/3}$.

-- The results of our numerical simulations are in contrast to the assumption normally found in literature \cite{GMS91,Ru86} , that Riemann waves in media with quadratic nonlinearity before breaking have a wave profile
of the form $x^{1/2}$, yielding a power law for the energy spectrum of $k^{-3}$. The difference can be seen clearly from Fig.\ref{f:8}.
\begin{figure}
\includegraphics[width=8cm]{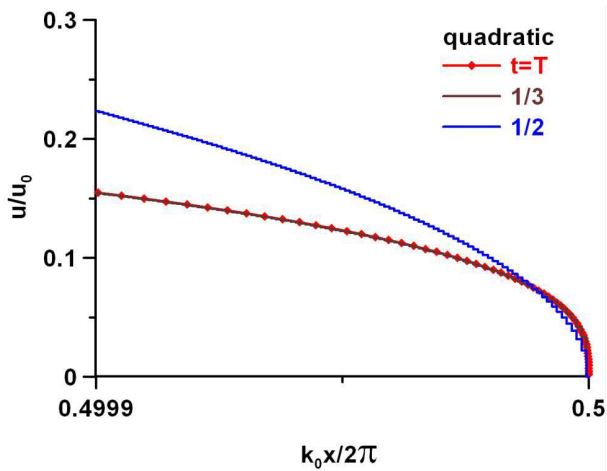}
\caption{\label{f:9} Color online. The shape of the Riemann wave in a quadratic medium (shown by black continuous line) and two singular
profiles: $x^{1/3}$ and $x^{1/2}$ shown by  red diamonds and  zigzag blue curve respectively.}
\end{figure}

-- We have demonstrated that the universal power law asymptotic of $k^{-8/3}$, inherited from the  simple wave equation with quadratic nonlinearity
\begin{equation}\label{Gardner}
u_t+ \a u  u_x = 0.
\end{equation}
"survives" for low-frequency range of harmonics in Fourier space with an additional dispersion term, yielding the KdV equation
\begin{equation}\label{Gardner-KdV}
u_t+ 6 u  u_x + u_{xxx}= 0 \nn
\end{equation}
and with an additional dissipation term, yielding the Burgers equation
\begin{equation}\label{Gardner-Bur}
u_t+ 6 u  u_x -0.1 u_{xx}= 0. \nn
\end{equation}
In both cases  the power law asymptotic in the low-frequency range changes in a similar way: the spectrum decay becomes slower.

A  really surprising  fact is that the universal power law asymptotic $k^{-8/3}$ obtained for  the breaking moment of time $T$ in  equation (\ref{Gardner}), is still observed both in the KdV and Burgers equation although breaking will not occur; moreover in the low frequency range it stays for  times $t \sim 2 T$ and more.

In the high-frequency range the difference between effects of dispersion and dissipation becomes visible.

With dissipation  taken into account, the well known power law asymptotic $k^{-2}$ of the Burgers equation is formed.

With  dispersion taken into account, a few soliton-like peaks in Fourier space are formed. The asymptotic of  their envelope has exponential form which was not studied in detail  yet. In our simulations  the number of solitons at the wave front was growing rapidly increasing the front width to the extent that the model equation is not applicable any more.

-- We wanted to  see how far the universal power asymptotic can be observed in other modifications of the simple wave equation. At our request, Karl Helfrich checked the shape of the energy spectrum in the results of his numerical simulations
with the reduced Ostrovsky equation
\be \label{Ostr-eq}
( u_t + u u_x )_x = \g u
\ee
with initial condition of the form (\ref{e5}). It turned out that the universal power asymptotic $k^{-8/3}$ is observed both in rotational and non-rotational cases, i.e. for arbitrary $\g$,  \cite{Ostr-Eq}.  The reduced Ostrovsky equation can be regarded as a modification of the Korteweg-de Vries
equation, in which the usual  dispersive term $u_{xxx}$ is replaced by the term $\g u$, which represents the effect of background rotation. The Ostrovsky equation is an important theoretical tool for studying surface and internal solitary waves in the atmosphere and ocean taking into account the Earth's rotation.

-- We found that in any weakly nonlinear medium which has been analyzed so far, non-dispersive or weakly dispersive, the time evolution of the energy spectrum goes through the same steps: an initially exponential  spectrum in the vicinity of the breaking point turns into a power law for high frequencies. This enables us to estimate from the spectrum measured how far a wave system is away from breaking (smooth form of wave, the beginning of collapse or  developed shock).  This method may be most useful in many problems of nonlinear  acoustics, physical oceanography, magnetohydrodynamics and  laser optics, see e.g. \cite{acoustic01,acoustic09,optic13}.

-- We have demonstrated - though not proven - that the power law asymptotic $k^{-8/3}$ of the energy spectrum of Riemann waves holds for the simple wave equation with arbitrary nonlinearity. The additional results we have found for modifications of the simple wave equation indicate, that the universal power law asymptotic  is a manifestation of some  fundamental properties which may be found in many wave systems and deserves further theoretical study. This is presently work in progress.

\textbf{{Acknowledgments.}} The authors acknowledge support by the
Austrian Science Foundation (FWF) under projects P22943 and
P24671. EP acknowledges also VolkswagenStiftung, RFBR grant
11-05-00216 and Federal Targeted Program УResearch and
educational personnel of innovation RussiaФ for 2009Ц2013.

\end{document}